\documentclass[12pt,a4paper]{article}
\usepackage[utf8]{inputenc}	% Para caracteres en español
\usepackage{amsmath,amsthm,amsfonts,amssymb,amscd}
\usepackage{multirow,booktabs}
\usepackage[table]{xcolor}
\usepackage{fullpage}
\usepackage{lastpage}
\usepackage{enumitem}
\usepackage{fancyhdr}
\usepackage{mathrsfs}
\usepackage{wrapfig}
\usepackage{setspace}
\usepackage{calc}
\usepackage{multicol}
\usepackage{cancel}
\usepackage[retainorgcmds]{IEEEtrantools}
\usepackage[margin=3cm]{geometry}
\usepackage{amsmath}

\usepackage{empheq}
\usepackage{framed}
\usepackage[most]{tcolorbox}
\usepackage{xcolor}
\colorlet{shadecolor}{orange!15}
\begin{document}

\def\ra{{\rightarrow}}
\def\a{{\alpha}}
\def\b{{\beta}}
\def\l{{\lambda}}
\def\eps{{\epsilon}}
\def\T{{\Theta}}
\def\t{{\theta}}
\def\co{{\cal O}}
\def\car{{\cal R}}
\def\caf{{\cal F}}
\def\cs{{\Theta_S}}
\def\pr{{\partial}}
\def\tri{{\triangle}}
\def\na{{\nabla }}
\def\S{{\Sigma}}
\def\s{{\sigma}}
\def\sp{\vspace{.08in}}
\def\hs{\hspace{.25in}}
\def\rs{\vspace{-.08in}}

\newcommand{\be}{\begin{equation}} \newcommand{\ee}{\end{equation}}
\newcommand{\bea}{\begin{eqnarray}}\newcommand{\eea}
{\end{eqnarray}}

\begin{titlepage}
\vspace*{\fill}
\begin{center}
      {\Huge  Geometric torsion, $4$-form, Riemann duals and Quintessence}\\[0.5cm]
      {\large R. Nitish and Supriya Kar}\\[0.4cm]
      {\large \slshape Department of Physics and Astrophysics,\\ University of Delhi 110007, India}\\[0.5cm]
     % \today
    \end{center}
 \begin{abstract}
        We revisit an emergent gravity scenario in $(4+1)$ dimensions underlying a propagating geometric torsion ${\cal H}_3$ with a renewed interest. We show that a pair-symmetric $4$th order curvature tensor is sourced by a two-form Neveu-Schwarz (NS)  in a $U(1)$ gauge theoretic formulation. Interestingly the new space-time curvature governs a torsion free geometry sourced by a two-form NS field and shares the properties of the Riemann tensor. On the other hand, a completely anti-symmetric $4$th order tensor in the formulation is shown to incorporate a dynamical geometric torsion correction and is argued to be identified with a non-perturbative correction. The four-form turns out to be $U(1)$ gauge invariant  underlying an onshell NS form. We show that an emergent gravity theory may elegantly be described with an axionic scalar presumably signifying a quintessence coupling to the Riemann type geometries. The curvatures are appropriately worked out to obtain a $d$$=$$12$ emergent {\it form} theory. Investigation reveals that a pair of $(M{\bar M})_{10}$-brane is created across an event horizon. We show that an emergent $M$ theory in a decoupling limit identifies with the bosonic sector of $N$$=$$1$ Supergravity in $d$$=$$11$.
\end{abstract}
    \vspace*{\fill}
\end{titlepage}

\section{Introduction} 
\noindent
Black holes are described in Einstein gravity and are governed by a metric dynamics on a Riemannian manifold. Interestingly, the stringy versions of these macroscopic black holes have been obtained in a low energy limit of a string effective action  \cite{gibbons-maeda-NPB,garfinkle,giddings-strominger}.

\sp
\noindent
In the past there have been attempts to use a gauge principle for the Riemannian geometry \cite{wilczek,Sharpe-torsion}. Interestingly the quantum effects to Einstein gravity may also be addressed non-perturbatively using the strong-weak coupling duality in ten dimensional superstring theories. A non-perturbative quantum effect is believed to be sourced by the compactified extra space dimension(s) to the stringy vacua. In particular the type IIA superstring theory in a strong coupling limit is known to incorporate an extra spatial dimension on $S^1$ and has been identified with an eleven dimensional non-perturbation (NP) $M$-theory. Generically $M$-theory has been shown to be identified with the stringy vacua in various dimensions. In a low energy limit $M$ theory is known to describe $N$$=$$1$ supergravity (SUGRA) in eleven dimensions \cite{cremmer-julia-scherk}. However a complete NP-theory is not fully been explored \cite{schwarz-PLB}.

\sp
\noindent
{Interestingly, it has been shown in the context of extended supergravity, one can express the curvatures in terms of torsion and its covariant derivatives thereby implying that torsion  has a geometrical origin \cite{dragon}. Also, the self duality of Born-Infeld action for  $3$-branes of Type IIB theory, underlying SL(2, Z) symmetry provide strong hints towards a $d$$=$$12$ generic form theory. The idea of duality in Type IIB superstring theory has been explored in the context of globally electric charged $(-1)$-branes(instantons) which are  dual to $7$-branes in $d$$=$$10$ and are known to describe spacetime wormhole underlying Riemannian geometry in Type IIB Supergravity theory \cite{gibbons96}.

\sp
\noindent
The formalism presented in the article deals only with the field theory underlying commutative spacetimes. However, an approach to emergent gravity theory underlying non-commutative (NC) spacetime has been explored  in the past\cite{seiberg-witten,Yuji-npb2001}. The presence of non-commutativity  leads to non-trivial modifications in the properties of black holes\cite{rizzo}. The idea provides a clue to explore a dynamical NS two-form in a modified torsion gravity originally sourced by a  gauge theory.  It has  been shown that gravity can emerge from a gauge theory in noncommutative (NC) spacetime. The NC gauge/gravity correspondence has been further explored for a constant two-form $B_{\mu\nu}$ background\cite{Miao1999}. For interesting details see refs \cite{yangmpla, yangijmpa, yangjhep, yangarxiv}. In addition some regular metric solutions, underlying non-commutative geometry, has been obtained  to describe de Sitter at small scale which approaches Riessner-Nordstr$\ddot{\rm o}$m geometry at large scale \cite{nicolini}. Interestingly  it has been shown that the  geometrical nature of gravity may be governed by a  torsion in an alternate formulation\cite{Niels-2018,Jose-arXiv-2019,TPS-arXiv2019}

\sp
\noindent
{Interestingly an effective field  description in $d$$=$$12$ has been proposed by considering supergravity theory in the same dimension. Such a theory describes Type IIA and IIB SUGRA upon compactification on $S^{1}$ and  is Poincar$\acute{\rm e}$ invariant when compactified on a torus\cite{kang}. Recently  it has been proposed that starting from a covariant formulation in $d$$=$$12$ and  it is possible to construct higher derivative effective field theory. Further it was shown to reduce to Type IIB SUGRA when compactified on a torus \cite{hamid}.}

\sp
\noindent
In the article we begin with $d$$=$$5$  Kalb-Ramond two-form $B_{\mu\nu}$ gauge theory.  In fact  a gauge theoretic torsion curvature $H_3$$=$$d^{\nabla}B_{2}$ on a $4$-brane has been explored to modify the derivative $\nabla_{\mu}$ to a covariant  derivative ${\cal D}_{\mu}$.    
We attempt to construct  $M$ theoretic action underlying a geometric torsion ${\cal H}_3$ dynamics in a $U(1)$  gauge theoretic formulation defined with ${\nabla}_{\mu}$ covariant derivative. In particular the ${\cal H}_3$$=$$d^{\cal D}B_{2}^{(NS)}$ has been shown to govern the dynamics of a Neveu-Schwarz (NS) two-form  on a fat brane in an emergent gravity  scenario \cite{abhishek-JHEP,abhishek-PRD,abhishek-NPB-P}. It was shown that the emergent theory is governed by a fourth order generalized curvature tensor ${\cal K}_{\mu\nu\lambda\rho}$ defined with two derivatives \cite{abhishek-JHEP}. Interestingly  a non-constant NS two-form background has been attempted to obtain a non-commutative D-brane in a different context \cite{Ho-prl2000}. 

\sp
\noindent
Furthermore  we explore the two derivative formulation underlying  a dynamical geometric torsion ${\cal H}_3$. In this paper  we explicitly realize a four-form ${\cal F}_4$$=$$d^{\nabla}{\cal H}_{3}$ field strength for an onshell NS two-form. In fact  the  NS-form gauge theory  has been shown to receive a dynamical geometric torsion correction which turns out to be a second derivative in NS-form. Together they define a $d$$=$$5$ NP-theory in the modified theory. In particular the dynamical correction ${\cal F}_4$ to the low energy Einstein vacuum has been argued to be sourced by an  instanton underlying an axionic scalar which  is  is believed to be non-perturbative in nature. 

\sp
\noindent
We show that our analysis generically reveals a fundamental theory in $d$$=$$12$ and  may represent a  {\it form} theory. It is shown that a dynamical quintessence field incorporates a NP-correction hidden to the torsion-less geometries. Analysis reveals that a fundamental building block of {\it form} theory, $i.e.$ an instanton, can be a potential candidate to describe the quintessence and hence dark energy in universe. At this point we may recall that  $(-1)$-branes are dual to $9$-branes in $D$$=$$10$ superstring theory.  In fact $9$-branes  are space filling and they have been viewed as a pair of a gravitational $(8\bar{8})$-brane pair configuration\cite{abhishek-JHEP}. 

\section{Higher form dynamics}
\subsection{Pair production scenarios} 
\noindent
We begin by briefly revisiting the Schwinger pair production mechanism \cite{schwinger}. The novel idea elegantly describes a $(e^+e^-)$ pair production underlying a QFT vacuum, by a photon. The mechanism is an emergent non-perturbative phenomenon and is believed to be instrumental to describe diverse quantum phenomenon in gravitation and cosmology. In fact this idea was explored to explain the Hawking radiation phenomenon \cite{hawking} where an incident photon generates a pair of charged particle/anti-particle at the event horizon of a black hole. 

\sp
\noindent
Furthermore  the idea was also applied to the open strings pair production \cite{bachas-porrati} and to investigate the production of a pair of $(D{\bar D})_9$ at the cosmological horizon \cite{majumdar-davis}. An extra spatial dimension on $S^1$ has been argued to unfold in $d$$=$$10$, type IIA superstring theory in a strong coupling and is believed to describe the $M$-theory in $d$$=$$11$.

\sp
\noindent
In the context an emergent theory of gravity on a stringy pair of $(3{\bar 3})$-brane by a Kalb-Ramond (KR)-form quanta on a $D_4$-brane has been formulated in a collaboration \cite{abhishek-JHEP,abhishek-PRD}. Importantly a generic vacuum geometry in general theory of relativity (GTR), $i.e.$ a Kerr family of black holes, has been obtained in a NP-decoupling limit which leads to the low energy (semi-classical) geometries \cite{sunita-NPB}.

\subsection{Emergent gravity in ${\mathbf 5}$D}
\noindent
Interestingly  the emergent gravity patches, underlying a non-linear higher form field quanta, were (spin) projected using a discrete matrix to describe an appropriate causal geometry in a low energy (GTR) limit \cite{priyabrat-EPJC}. The emergent gravity patches have formally been viewed as a lower dimensional $D_p$-brane correction to a low energy string vacuum \cite{priyabrat-EPJC}. Since a $D_p$-brane carries Ramond-Ramond (RR) charge, the correction turns out to be non-perturbative \cite{polchinski}. 

\sp
\noindent
It was argued that an emergent pair breaks the supersymmetry due to the exchange 
of closed string modes in between the vacuum created gravitational pair of $(3{\bar 3})$-brane across the cosmological horizon \cite{abhishek-JHEP}. Thus, the emergent gravitational pair is described on a fat brane underlying a massive NS field dynamics in a first order theory. In fact, a quintessence scalar field acts as an extra (fifth) hidden dimension inbetween an emergent 
gravitational brane pair. It determines the thickness of a fat brane. 

\sp
\noindent
Alternately  a quintessence degree of freedom, underlying the closed string exchange between the gravitational pair, is absorbed by a $D_3$-brane to make it a gravitational fat $3$-brane. In the context a vacuum ($T_{\mu\nu}$$=$$0$) refers to a background black hole defined with an open string metric on a $D_4$-brane.
\bea
{\tilde{G}}_{\mu\nu}=\big ( g_{\mu\nu}-B_{\mu}^{(NS)\lambda}B^{(NS)}_{\lambda\nu}\big )\ ,
\eea
where $g_{\mu\nu}$ is a metric underlying vanishing Riemannian curvature.  Thus  the open string metric is sourced by a constant (global mode)  NS field \cite{seiberg-witten} in the string bulk. In addition the Poincar$\acute{\rm e}$ duality in the five dimensional world-volume gauge theory ensures a two-form  KR dynamics on a $D_4$-brane. Thus the non-linear $U(1)$ gauge dynamics of a $D_4$-brane in presence of a background metric may be envisaged with a constant NS field and a local KR field. It is described by
\be
S={{-1}\over{(8\pi^3g_s){\alpha'}^{3/2}}}\int d^5x {\sqrt{-{\tilde G}^{({\rm NS})}}}\ H_{\mu\nu\lambda}H^{\mu\nu\lambda}\ ,\label{p-1}
\ee
where ${\tilde{G}}=$$\det{\tilde{G}}_{\mu\nu},\ H_{\mu\nu\lambda}$$=$$\big ( \nabla_{\mu}B_{\nu\lambda}$$+{\rm cyclic}\big )$, $g_{s}$ is the string coupling and $\alpha'=$ string length parameter. Generically  the idea is to explore a higher-form $U(1)$ gauge theory and hence its geometric effects  on an appropriate $D_p$-brane. In fact a higher-form gauge theory is non-linear and becomes sensible in higher dimensions. It may provide a clue to unfold a NP-theory of gravity in higher dimensions such as $M$-theory in eleven dimensions \cite{schwarz-PLB}. A priori for simplicity one may  begin with a KR two-form gauge theory on a $D_4$-brane in a ten dimensional type IIA superstring theory. Nevertheless a five dimensional construction is a minimal extension to the GTR and is believed to be a fundamental unit. Needless to mention that GTR is an elegant construction by Einstein in $d$$=$$4$ defined with a vacuum solution whereas BTZ black hole in $d$$=$$3$ is a brilliant construction for a non vacuum solution \cite{BTZ}.  As a bonus our formulation $d$$=$$5$  takes  account for the quint-essential cosmology. Quintessence has been explored in the context  \cite{priyabrat-EPJC}. It is believed to be a candidate for  the observed accelerated rate of expansion of universe and the origin of dark energy.

\subsection{NS field dynamics}
\noindent
An emergent fat brane evolves with a dynamical NS field which is obtained at the expense of the KR field dynamics on a $D_4$-brane. Thus, a covariantly constant NS field, $i.e.\ \nabla_{\lambda}B_{\mu\nu}^{(NS)}=0$, on a $D_4$-brane becomes dynamical on a fat brane. A geometric torsion ${\cal H}_3$ is
constructed using a modified derivative:
\bea
&&{\cal H}_{\mu\nu\lambda}={\cal D}_{\mu}B^{(NS)}_{\nu\lambda}+{\cal D}_{\nu}B^{(NS)}_{\lambda\mu}+{\cal D}_{\lambda}B^{(NS)}_{\mu\nu}
\ ,\qquad\; {}\nonumber\\
{\rm where}\qquad &&{\cal D}_{\lambda}B^{(NS)}_{\mu\nu}={1\over2}\left ({H_{\lambda\mu}{}}^{\rho}B^{(NS)}_{\nu\rho}+ {H_{\lambda\nu}{}}^{\rho}B^{(NS)}_{\rho\mu}\right ).\label{gtorsion-1}
\eea
A constant NS field background in a closed string, becomes a nontrivial geometric torsion at the expense of the the KR field dynamics, $i.e.\ {\cal D}_{\lambda}B_{\mu\nu}=0$. It implies that the modified derivative ${\cal D}_{\mu}$ is uniquely fixed. In fact, the absorption of KR field quanta, in a background black hole, has given birth to an emergent quantum theory of gravitation on a stringy pair of $(3{\bar 3})$-brane across an event horizon \cite{abhishek-PRD}. Interestingly, a plausible NP-theory of gravity underlying an emergent scenario enforces an iterative correction, where the KR-form dynamics is replaced by the NS-form, $i.e.\ H_3\rightarrow {\cal H}_3$. It leads to an exact description with a $B^{(NS)}_2$ perturbation:
\be
{\cal H}_{\mu\nu\lambda}=H_{\mu\nu\rho}B_{\lambda}^{(NS)\rho}+ H_{\mu\nu\alpha}B_{\rho}^{(NS)\alpha} B_{\lambda}^{(NS)\rho}+\dots\label{gtorsion-2}
\ee
For an infinitesimal perturbation, the geometric torsion: ${\cal H}_3\rightarrow H_3$. A priori an exact perturbation NS field theory may equivalently be described by a NP-theory of gravity whose gauge potential shall be shown to be sourced by a dynamical NS field. 

\sp
\noindent
In fact there are three distinct dynamical aspects of the same description in five dimensions and they are: (i) the KR perturbation theory underlying a gauge group $U(1)$ on a $D_4$-brane and is defined with a covariant derivative $\nabla_{\mu}$, (ii) the NS perturbation theory of emergent semi-classical gravity on a fat brane underlying the gauge group $U(1)$ and is defined with a modified covariant derivative ${\cal D}_{\mu}$, (iii) the geometric torsion non-perturbation theory of emergent quantum gravity on a vacuum created $(3{\bar 3})$-brane underlying an enhanced gauge group $U(1)\otimes U(1)_{\rm NP}$. An additional $U(1)$ in a NP-theory is only sensible in second order while the KR and NS perturbation gauge theories are first order formulations. Both, the original and the emergent, perturbation theories are respectively described by the $U(1)$ gauge invariant field strengths $H_3$ in eq(\ref{p-1}) and ${\cal H}_3$ in eq(\ref{gtorsion-1}). However, the NP-theory breaks the gauge invariance perturbatively which is evident from the field strength expression in eq(\ref{gtorsion-2}). The perturbative gauge invariance has been realized \cite{abhishek-JHEP,priyabrat-EPJC} in presence of a symmetric fluctuation: $f_{\mu\nu}={\bar{\cal H}}_{\mu\alpha\beta}{{\cal H}^{\alpha\beta}{}}_{\nu}$, where ${\bar{\cal H}}_3=(2\pi\alpha'){\cal H}_3$. Thus a dynamical NS field modifies the constant background metric $G_{\mu\nu}^{(NS)}$ into a dynamical metric on an emergent fat brane. The modified metric is given by
\be
G_{\mu\nu}=\Big ( g_{\mu\nu}-B_{\mu}^{(NS)\lambda}B^{(NS)}_{\lambda\nu} + {\bar{\cal H}}_{\mu\lambda\rho}{{\cal H}^{\lambda\rho}}_{\nu}\Big )\ . \label{gauge-metric}
\ee
Firstly, it shows that a geometric torsion incorporates an intrinsic angular momentum in an emergent vacuum. Thus an emergent gravity naturally governs the Kerr black hole as a vacuum geometry \cite{sunita-NPB}. 

\sp
\noindent
Secondly, the emergent metric (\ref{gauge-metric}) dynamics does not emerge in first order perturbation theory and therefore clearly requires at least a second order formulation. Most importantly, the gauge invariance shall shown to be restored in a second order for an onshell NS field in the paper. As a result, the geometric torsion ${\cal H}_3$ is perceived as a gauge potential underlying a NP-dynamical correction to the perturbative NS field dynamics in first order and hence the complete NP-theory is governed in $1.5$ order by ${\cal H}_3$.

\sp
\noindent
At a first sight it may imply that the modified metric dynamics is primarily governed by a propagating ${\cal H}_3$. However a geometric torsion theory, being a quantum description, cannot be identified with a metric tensor theory. This is due to a fact that the (pseudo) Riemannian geometry is sourced by a metric tensor (GTR). The apparent puzzle is resolved with an emergent metric which is a semi-classical phenomenon as is realized in a first order perturbation theory of a NS field. The complete NP-theory in $1.5$ order is an emergent quantum gravity description underlying a geometric torsion for the onshell NS field. Thus, the notion of a dynamical metric is not sensible due to the prevailing geometric torsion dynamics in a NP-theory of quantum gravity in $d$$=$$5$. A propagating torsion is known to break the space-time continuum and hence there are no closed geodesics.

\sp
\noindent
In the paper, we explicitly work out a NP-dynamical correction in a $d$$=$$5$ emergent scenario which in turn, is a key to unfold an emergent $M$-theory in eleven dimensions underlying a {\it fundamental} ({\it form}) theory in twelve dimensions. It has been shown that the GTR emerges as non-perturbative phenomenon on a stable gravitational pair of $(3{\bar 3})$-brane \cite{abhishek-JHEP,abhishek-PRD,abhishek-NPB-P}. A fact that GTR is a fundamental unit, or a $3$-brane in the formulation, allows one to believe in a twelve dimensional {\it fundamental} theory which is worked out in this paper. In addition, a coupling between the Riemann tensor and a NP-dynamics has been realized which signals the intrinsic role played by a geometric torsion in the GTR. 
\section{New curvatures in modified gravity}
\subsection{Emergent curvature tensors $\big ({{{\cal H}_{\mu\nu}{}}^{\lambda}},\ {{\cal K}_{\mu\nu\lambda\rho}},\ {{\cal L}_{\mu\nu\lambda\rho}}\big )$}
\noindent
The commutator of modified derivatives have been worked out to obtain the generic curvature tensors \cite{abhishek-JHEP}. It was shown that 
an emergent curvature describes a vacuum created gravitational pair of $(3{\bar 3})$-brane \cite{abhishek-PRD}. The commutators yield:
\bea
 &&\left [ {\cal D}_{\mu}\ ,\ {\cal D}_{\nu}\right ]\psi\ =\  -2\ {{\cal H}_{\mu\nu}{}}^{\lambda}\ {\cal D}_{\lambda}\psi\nonumber\\
 {\rm and} \qquad &&\left [ {\cal D}_{\mu}\ ,\ {\cal D}_{\nu}\right ]A_{\lambda}\ =\  \left ({{\cal R}_{\mu\nu\lambda}}^{\rho}+{{\cal K}_{\mu\nu\lambda}{}}^{\rho}
\ +\ {{\cal L}_{\mu\nu\lambda}}^{\rho}\right )A_{\rho} +{{\cal H}_{\mu\nu}}^{\rho}{\cal D}_{\rho}A_{\lambda}\ ,\label{new-curvature}
\eea
where ${\cal H}_3$ ensures a NS field dynamics in an emergent first order perturbation theory. 
It provokes thought to believe that the quanta of NS field in superstring theory can govern an emergent graviton. 

\sp
\noindent
Furthermore, the coupling of ${\cal H}_3$ to the {\it dynamics} ${\cal D}_{\lambda}\psi$ rather than the field $\psi$ in eq(\ref{new-curvature}) is a new phenomenon. A non-zero ${\cal H}_3$ signifies a difference in energy between (emergent) quantum gravity and classical (Einstein) gravity. In fact, ${\cal D}_{\mu}\psi\neq 0$, $i.e.$ the dynamical scalar, is known to be pivotal to incorporate a non-perturbative (quantum) correction to the GTR \cite{priyabrat-EPJC}. Thus, a dynamical correction to the GTR plays a significant role in quantum cosmology and is indeed a powerful tool to explain the accelerated rate of expansion of the universe. The inherent (quantum) dynamics of the scalar field makes the emergent gravity a natural candidate to describe the quintessence which is believed to govern the dark energy in the universe.

\sp
\noindent
The fourth order reducible emergent curvature tensors (\ref{new-curvature}) play a significant role in a second order non-perturbation theory of emergent gravity. They are believed to be potential keys to a complete non-perturbation theory such as $M$ theory.The curvature ${\cal R}_{\mu\nu\lambda\rho}$ is the well known Riemann curvature tensor, sourced by a metric field in $d$$=$$5$. It's very presence in the commutator implies the fact that the one-form $A_{\mu}$, on which the modified derivatives ${\cal D}_{\mu}$ act may be non-linear in origin. The second emergent curvature,
\begin{equation}
{{\cal K}_{\mu\nu\lambda}}^{\rho} = \frac{1}{2}\partial_{\mu}{{\cal H}_{\nu\lambda}}^{\rho} - \frac{1}{2}\partial_{\nu}{{\cal H}_{\mu\lambda}}^{\rho} + \frac{1}{4}{{\cal H}_{\mu\lambda}}^{\sigma}{{\cal H}_{\nu\sigma}}^{\rho} - \frac{1}{4}{{\cal H}_{\nu\lambda}}^{\sigma}{{\cal H}_{\mu\sigma}}^{\rho} 
\end{equation}
is expressed as pair symmetric $(S)$ and  pair non-symmetric $({\tilde A})$ under an interchange of first and second pair of indices, $i.e.\ {\cal K}^{(S)}_{\mu\nu\lambda\rho}+{\cal K}^{({\tilde A})}_{\mu\nu\lambda\rho}$. Explicitly all three emergent curvatures are:
\bea
{\cal R^{\lambda}}_{\mu\nu\rho}&=& \partial_{\rho}\Gamma^{\lambda}_{\mu\nu}-\partial_{\nu}\Gamma^{\lambda}_{\mu\rho}+\Gamma^{\sigma}_{\mu\nu}\Gamma^{\lambda}_{\rho\sigma}-\Gamma^{\sigma}_{\mu\rho}\Gamma^{\lambda}_{\nu\sigma}\ , \nonumber \\
{{\cal L}_{\mu\nu\lambda}}^{\rho}&=&\frac{1}{2} \Big (\Gamma^{\rho}_{\mu\sigma} {{\cal H}^{\sigma}}_{\nu\lambda}+\Gamma^{\sigma}_{\nu\lambda} {{\cal H}^{\rho}}_{\mu\sigma} - \Gamma^{\sigma}_{\mu\lambda} {{\cal H}^{\rho}}_{\nu\sigma} -\Gamma^{\rho}_{\nu\sigma} {{\cal H}^{\sigma}}_{\mu\lambda} \Big )\nonumber \\
%\eea
%\bea
{\rm and} \qquad \qquad \quad {\cal K}^{(S)}_{\mu\nu\lambda\rho}&=& \frac{1}{4}{{\cal H}_{\mu\lambda}}^{\sigma}{\cal H}_{\nu\sigma\rho}-\frac{1}{4}{{\cal H}_{\nu\lambda}}^{\sigma}
{\cal H}_{\mu\sigma\rho} \ , \nonumber\\
 \qquad \qquad \qquad \quad{\cal K}^{({\tilde A})}_{\mu\nu\lambda\rho}&=&\frac{1}{2}\partial_{\mu}{\cal H}_{\nu\lambda\rho} -\frac{1}{2}\partial_{\nu} {\cal H}_{\mu\lambda\rho}\ ,\label{g-curvature}
\eea
where  $\Gamma^{\rho}_{\mu\sigma}$ denote the Christoffel connections sourced by the background metric ${\tilde G}_{\mu\nu}$. Thus ${{\cal L}_{\mu\nu\lambda}}^{\rho}$ describes a coupling between the GTR and an emergent gravity. It ensures a non-propagating geometric torsion and hence a torsion free geometry. An irreducible tensor is found to yield \cite{abhishek-JHEP}:
\be
\delta^{\nu}_{\rho}{{\cal L}_{\mu\nu\lambda}}^{\rho}={\cal L}_{\mu\lambda}=\frac{1}{2}\Big (\Gamma^{\nu}_{\mu\rho} {{\cal H}^{\rho}}_{\nu\lambda}+\Gamma^{\rho}_{\nu\lambda} {{\cal H}^{\nu}}_{\mu\rho}- \Gamma^{\nu}_{\nu\rho} {{\cal H}^{\rho}}_{\mu\lambda}\Big )=-{\cal L}_{\lambda\mu}\ .\label{gauge-65}
\ee
The Lorentz scalar $\big ({\cal L}_{\lambda\mu}{\cal L}^{\lambda\mu}\big )$ involves quadratic power in geometric torsion and Christoffel connection(s). Since the  perturbation parameter is small, quadratic and higher powers become insignificant. A priori, it may imply that an emergent gravity in an all order perturbation theory is defined in a decoupling limit of two distinct connections. However, a non-perturbation emergent theory incorporates the coupling of connections defined with the Lorentz scalar 
$\big ({\cal L}_{\lambda\mu}{\cal L}^{\lambda\mu}\big )$.

\sp
\noindent
Furthermore  the emergent curvature ${{\cal K}_{\mu\nu\lambda}}^{\rho}$ in terms of covariant derivative $\nabla_{\mu}$ becomes
\begin{eqnarray}
\nabla_{\mu}{{\cal H}_{\nu\lambda}}^{\rho}- \nabla_{\nu}{{\cal H}_{\mu\lambda}}^{\rho}&= \partial_{\mu}{{\cal H}_{\nu\lambda}}^{\rho}- \partial_{\nu}{{\cal H}_{\mu\lambda}}^{\rho}+2 {{\cal L}_{\mu\nu\lambda}}^{\rho}\nonumber\\
{\rm or}\quad  \frac{1}{2}\partial_{\mu}{{\cal H}_{\nu\lambda}}^{\rho}- \frac{1}{2}\partial_{\nu}{{\cal H}_{\mu\lambda}}^{\rho} &= \frac{1}{2}\nabla_{\mu}{{\cal H}_{\nu\lambda}}^{\rho}- \frac{1}{2}\nabla_{\nu}{{\cal H}_{\mu\lambda}}^{\rho}-{{\cal L}_{\mu\nu\lambda}}^{\rho} \ ,
\end{eqnarray}
where the emergent curvature ${{\cal L}_{\mu\nu\lambda\rho}}^{\rho}$ is defined by equation (\ref{g-curvature}). Thus, the fourth order emergent curvature can be expressed as
\begin{equation}
{\cal K}_{\mu\nu\lambda\rho}={\cal K}_{\mu\nu\lambda\rho}^{(s)}+ \Big (\frac{1}{2}\nabla_{\mu}{\cal H}_{\nu\lambda\rho}- \frac{1}{2}\nabla_{\nu}{\cal H}_{\mu\lambda\rho}\Big )-{\cal L}_{\mu\nu\lambda\rho}\ ,
\end{equation}
where the second term in the bracket in the emergent curvature ${\cal K}_{\mu\nu\lambda\rho}^{(\tilde{A})}$ in equation(\ref{g-curvature}). The only difference is that, here the operating derivative is $\nabla_{\mu}$ instead of ordinary derivative $\partial_{\mu}$.
The commutator (\ref{new-curvature}), simplifies to
\begin{equation}\label{commutator-01}
 [{\cal D}_{\mu} , {\cal D}_{\nu}]A_{\lambda} =\Big ({\cal R}_{\mu\nu\lambda\rho}+{\cal K}_{\mu\nu\lambda\rho}^{(s)}+ \frac{1}{2}\nabla_{\mu}{\cal H}_{\nu\lambda\rho}- \frac{1}{2}\nabla_{\nu}{\cal H}_{\mu\lambda\rho} \Big )A^{\rho}+{{\cal H}_{\mu\nu}}^{\rho}{\cal D}_{\rho}A_{\lambda}\ .
\end{equation}  
\sp
\noindent
The emergent curvature tensor ${\cal K}^{(S)}_{\mu\nu\lambda\rho}$ of order four (\ref{g-curvature}) shares all the properties of a Riemann tensor under the interchange of indices within a pair and with pairs. In addition, the pair-symmetric curvature is sourced by a NS field propagation in an emergent perturbation theory in first order. As explained, it does not depend on the propagation of a ${\cal H}_3$ which is a second order phenomenon. Thus, it describes a torsion free geometry. Thus the pair-symmetric tensor may be identified with the Riemann type tensor ${\cal R}_{\mu\nu\lambda\rho}$.  The emergent curvature ${\cal K}_{\mu\nu\lambda\rho}^{(\tilde{A})}$ in eq(\ref{g-curvature}), at the expense of another emergent curvature ${\cal L}_{\mu\nu\lambda\rho}$, may be expressed as,
\begin{eqnarray}
{\cal K}_{\mu\nu\lambda\rho}^{(\tilde{A})} &=&  \frac{1}{2}\nabla_{\mu}{\cal H}_{\nu\lambda\rho}- \frac{1}{2}\nabla_{\nu}{\cal H}_{\mu\lambda\rho}\\
 {\rm or} \qquad{\cal K}_{\mu\nu\lambda\rho}^{(\tilde{A})}&=& \frac{1}{2} \Big (\nabla_{\mu}{{\cal H}_{\nu\lambda\rho}} - \nabla_{\rho}{\cal H}_{\mu\nu\lambda}+\nabla_{\lambda}{\cal H}_{\rho\mu\nu}-\nabla_{\nu}{{\cal H}_{\lambda\rho\mu}} \Big )\nonumber \\
&+& \frac{1}{2} \nabla_{\rho}{\cal H}_{\mu\nu\lambda} -\frac{1}{2}\nabla_{\lambda}{\cal H}_{\rho\mu\nu} \ , \\
&=&  \frac{1}{2} \Big (\nabla_{\mu}{{\cal H}_{\nu\lambda\rho}} - \nabla_{\rho}{\cal H}_{\mu\nu\lambda}+\nabla_{\lambda}{\cal H}_{\rho\mu\nu}-\nabla_{\nu}{{\cal H}_{\lambda\rho\mu}} \Big ) + {\cal K}^{(A)}_{\mu\nu\lambda\rho}\ ,
\end{eqnarray}
where in the second term, we have added and subtracted two covariant derivative terms. The commutator (\ref{commutator-01}), using above equations, can be put into much more elegant form,
\be
 [{\cal D}_{\mu} , {\cal D}_{\nu}]A_{\lambda} =\Big ({\cal R}_{\mu\nu\lambda\rho}+{\cal K}_{\mu\nu\lambda\rho}^{(S)}+ {1\over{\sqrt{8 \pi \alpha'}}}{\cal F}_{\mu\nu\lambda\rho} \Big )A^{\rho}+{\cal K}_{\mu\nu\lambda\rho}^{(A')}A^{\rho}+{{\cal H}_{\mu\nu}}^{\rho}{\cal D}_{\rho}A_{\lambda} \ .
\ee
The emergent curvatures ${\cal F}_{\mu\nu\lambda\rho}$ and ${\cal K}_{\mu\nu\lambda\rho}^{(A')}$ are completely antisymmetric and pair asymmetric tensors respectively.

\vspace{.08in}
\noindent
A fact that the pair symmetric generalized curvature ${\cal K}^{(S)}_{\mu\nu\lambda\rho}$, sourced by a dynamical two-form, precisely shares the tensor properties of the Riemann tensor is remarkable. Interestingly, the curvature is an invariant notion and its source, \emph{i.e.} a (gauge) potential is not. The observation may provoke thought to believe that a dynamical two-form may incorporate a Riemannian geometry notionally. Needless to mention that the pair symmetric ${\cal K}^{(S)}_{\mu\nu\lambda\rho}$ is free from torsion ${\cal H}_{\mu\nu\lambda}$ dynamics and further ensures its consistency with the Riemannian geometry. On the other hand, the remaining part of the generalized curvature, ${\cal K}_{\mu\nu\lambda\rho}^{(\tilde{A})}$ ensures a propagating torsion ${\cal H}_{\mu\nu\lambda}$ which in turn has been identified with a four-form ${\cal F}_{\mu\nu\lambda\rho}$ in the formulation. Thus, the ${\cal F}_{4}$ is believed to incorporate a dynamical correction to the Riemann like ${\cal K}_{\mu\nu\lambda\rho}^{(S)}$ in the formulation. It may lead to an emergent quantum gravity description. In the context we may briefly revisit an interesting feature of an extremal Reissner-Nordstr$\ddot{o}$m (RN) black hole in Newtonian limit. The shrinking event horizon of the RN black hole ensures the decoupling of the non-linear modes of Einstein gravity and leads to a stable extremal RN black hole where mass is identified with the electric charge. Interestingly, the extremal limit may be identified with an equipotential underlying two distinct elegant formulations such as the Newtonian gravity and the Electro-Magnetic theory. The attractive gravitational force between two massive and electrically charged objects in the limit is precisely cancelled by the repulsive coulombic force. Thus, an extremal RN black hole may provide a clue to explore some quantum phenomenon and is likely to bring-in a new shade to quantum gravity. The essentil idea contained in the analysis may help to explain a possibility of cancellation of the pair symmetric ${\cal K}_{\mu\nu\lambda\rho}^{(S)}$ and Riemann curvatures. This in turn would govern a propagating (geometric) torsion whose field strength is defined as ${\cal F}_{\mu\nu\lambda\rho}=\nabla_{\mu}{\cal H}_{\nu\lambda\rho}+$ cyclic. 

\subsection{Dynamical correction}
\noindent
The ${\cal H}_3$ dynamics may be re-expressed with a pair anti-symmetric $(A)$ tensor:
\be
\qquad \qquad {\cal K}^{({\tilde A})}_{\mu\nu\lambda\rho}\ =\ {1\over{2\sqrt{2\pi\alpha'}}}\ {\cal F}_{\mu\nu\lambda\rho} \ +\ {\cal K}^{(A')}_{\mu\nu\lambda\rho}\ ,\label{gauge-7777}
\ee
where ${\cal F}_{\mu\nu\lambda\rho}=4{\sqrt{2\pi\alpha'}}\ {\nabla}_{[\mu}{\cal H}_{\nu\lambda\rho]}$.
The second term is described by an irreducible curvature: 
\be
\qquad \qquad {\cal K}_{\nu\rho}^{(A')}=g^{\mu\lambda}{\cal K}_{\mu\nu\lambda\rho}^{(A')}=-{\cal K}_{\rho\nu}^{(A')}=\frac{1}{2}\nabla_{\mu}{{\cal H}^{\mu}}_{\rho\nu}
\ .\label{gauge-77}
\ee
Thus, onshell NS-form implies that the two-form ${\cal K}^{(A')}_{\mu\nu}$ becomes trivial in the action. Then the Lorentz scalar:
\be
\qquad \qquad{\cal K}^{({\tilde A})}_{\mu\nu\lambda\rho} {\cal K}_{(\tilde{A})}^{\mu\nu\lambda\rho}\ =\ {1\over{8\pi\alpha'}} {\cal F}_{\mu\nu\lambda\rho}{\cal F}^{\mu\nu\lambda\rho}
\ .\label{gauge-792}
\ee
Irreducibility of the pair-symmetric tensor (\ref{g-curvature}) is worked out by taking its trace and we to obtain: ${\cal K}^{(S)}_{\mu\nu}=-\frac{1}{4}\big ({\cal H}_{\mu\rho\lambda}{{\cal H}^{\rho\lambda}{}}_{\nu}\big )$ and ${\cal K}^{(S)}=-\frac{1}{4}{\cal H}_3^2$. They describe torsion free geometries and may 
formally be identified with the Ricci type tensors: ${\cal R}_{\mu\nu}$ and ${\cal R}$. They lead to an emergent metric in a first order perturbation gauge theory which is precisely governed by a NS field dynamics on a fat brane. However, a dynamical correction (\ref{gauge-792}), underlying a propagating geometric torsion, becomes sensible with a NP-correction in a second order formulation. Thus, an emergent theory of NP-gravity may completely be perceived in $1.5$ order formulation with two irreducible curvature tensors ${\cal K}^{(S)}$ and ${\cal F}_4$.  

\subsection{Topological couplings}
\noindent
Interestingly the dynamical aspects of NP-gravity may seen to be supplemented with topological couplings. 
A priori, there are two topological couplings between a constant NS-form and a dynamical KR-form in the world-volume gauge theory on a $D_4$-brane. Both of them turn out to be a total divergence in the $d$$=$$5$ bulk (B), though they contribute significantly to the boundary (BD) theory. They are given by
\bea
{1\over{{\kappa'}^4}}\int_{\rm B} H_3\wedge\Big (B_2^{(NS)}+{\bar F}_2\Big )={1\over{\kappa^4}}\int_{\rm BD} B_2^{(KR)}\wedge {\bar{\cal F}}_2, \label{topological-1}
\eea
where $\kappa'$ and $\kappa$ are respectable couplings in $d$$=$$5$ and $d$$=$$4$. They possess a dimension of length.
Remarkably, the boundary action can be identified with the $BF$-topological theory. In an emergent scenario, the topological 
coupling is re-expressed as:
\be
{1\over{\kappa'}^4}\int_{\rm B} B_2^{(KR)}\wedge {\cal H}_3\ =\ {1\over{\kappa}^4}\int_{\rm BD} B_2^{(KR)}\wedge B_2^{(NS)}\ .\label{topological-2}
\ee
In addition the NS and RR couplings are known to incorporate a notion of branes within a brane. This novel idea has been pivotal to the formulation of two independent $d$$=$$10$ Matrix theories with: (i) $D$-particle as a building block for type IIA superstring \cite{BFSS-PRD} and (ii) $D$-instanton for type IIB superstring \cite{IKKT-NPB}.

\sp
\noindent
Though a topological coupling is trivial in the bulk, it regains significance in the boundary theory. It is given by 
\be
{1\over{\kappa'}^3}\int_{\rm B}{\cal F}_4\wedge {\cal F}_1={1\over{\kappa}^2}\int_{\rm BD} {\cal H}_3\wedge d\psi
={{\Lambda}\over{\kappa}^3}\int_{\rm BD} \psi
\ ,\label{topological-3}
\ee
where ${\cal H}_3$ turns out to be a $U(1)$ gauge potential and $\psi$ denotes an axion coupling. Primarily the broken $U(1)$ gauge invariance of the 3-form geometric torsion (\ref{gtorsion-2}) under the NS-form transformation identifies ${\cal H}_3$ with a gauge potential whose gauge invariant field strength is ${\cal F}_4$ and is obtained in eq(\ref{gauge-7777}). Interestingly, a gauge invariant four-form has been argued to be a candidate to tackle the cosmological constant problem \cite{Hawking-PLB-1998}.   Non-zero $\Lambda=\big ({\cal E}^{\mu\nu\lambda\rho}\ {\cal F}_{\mu\nu\lambda\rho}\big )$ may be identified with a cosmological constant in the boundary theory. Its toplogical coupling to an axionic scalar field in $d$$=$$4$ boundary theory is remarkable. It may imply that a topological coupling of an axion in $d$$=$$4$ plays an important role to describe a hidden fifth essence, $i.e.$ quintessence.

\subsection{Gravity duals}
\noindent
Now we revisit the construction for left and right duals of the Riemann tensor with a renewed interest using the $4$-form in the NP-formulation. The Riemann duals may be envisaged to absorb the dynamical axion and hence the instanton quantum effect in a semi-classical limit. Alternately, the quintessence cosmology, $i.e.$ the accelerated rate of expansion of universe, may well be described with the renewed construction for the Riemann duals and may serve as a viable source for dark energy.

\sp
\noindent
The pair-symmetric $(S)$ and the pair non-symmetric  $({\tilde A})$ emergent curvature 
tensors  of order four (\ref{g-curvature}) are worked out for their irreducibility to yield a curvature scalar ${\cal K}^{(S)}$ $\equiv$ ${\cal R}$ and ${\cal F}_4$ respectively. The four-form incorporates a dynamical correction obtained via eqs(\ref{gauge-7777})-(\ref{gauge-792}) in second order underlying an onshell NS-form in a first order gauge theory. The field strength ${\cal F}_4$ further ensures a propagating ${\cal H}_3$ and may seen to be governed by a lower dimensional $D$-brane correction \cite{priyabrat-EPJC}. Since D-branes are known to be non-perturbative objects  \cite{polchinski}, the effective action underlying an emergent quantum gravity on a gravitational pair of $(3{\bar 3})$-brane is believed to be non-perturbative. In fact, the emergent gravity on a pair of brane and anti-brane in the efffective gauge theory has been argued to be sourced by the absorption of NS-form quanta at the event horizon of a background stringy black hole \cite{abhishek-JHEP}. Interestingly, it generalizes the Schwinger pair production mechanism in quantum field theory \cite{schwinger} which is established as a non-perturbative phenomenon. The emergent gravity action may be given by
\be
S_{EG}={1\over{{\kappa'}^3}}\int d^5x {\sqrt{-g}}\; \Big ( {\cal R}-{1\over{2\cdot 4!}}\ {\cal F}_4^2\Big )\ ,\label{gauge-70}
\ee
where the $g_{\mu\nu}$ signature is $(-,+,+,+,+)$ and the string slope parameter $\alpha'$ is used to re-express ${\kappa'}^2=(8\pi\alpha')$. The first term governs a torsion free geometry. Importantly the second term incorporates a dynamical geometric torsion correction non-perturbatively to the Einstein vacuum. Furthermore, the Poincar$\acute{e}$ dual of $4$-form ensures the presence of a dynamical axionic scalar in the string-brane formulation for a NP-theory. In fact the axion is believed to source a instanton \cite{priyabrat-EPJC} which ensures an instantaneous quantum correction to the Einstein gravity. The instanton correction further reconfirms its non-perturbation nature and may be viewed as a source to the dark energy in universe.

\sp
\noindent
Remarkably, the emergent gravity action in a decoupling limit formally can be identified with the bosonic sector of $d$$=$$11$ SUGRA  \cite{cremmer-julia-scherk}. Thus a dynamical geometric torsion ${\cal H}_3$ potential may provide a clue to unfold a non-perturbation theory of quantum gravity. The emergent gravity action is re-expressed as: 
\be
S_{EG}={1\over{{\kappa'}^3}}\int_{\rm B} {\sqrt{-g}}\, {\cal R} -{1\over{96 {\kappa'}^3}}\int_{\rm B}{\cal F}_4 \wedge {}^{\star}d\psi
+{{\Lambda}\over{{\kappa}^3}}\int_{\rm BD} \psi\label{Egravity-dual}
\ee
The equivalence between (\ref{gauge-70}) and (\ref{Egravity-dual}) reconfirms a topological coupling and hence ${\cal F}_4$ in eq(\ref{gauge-7777}) modifies to: ${\sqrt{2\pi\alpha'}}\big ( d{\cal H}_3-{\cal H}_3\wedge d\psi\big )$. A dual action ensures a dynamical axionic scalar. The metric signature reverses to $(+,-,-,-,-)$ in the dual scenario due to the odd dimensional space-time. The dual action is given by
\be
S_{EG}^{Dual}\equiv\int_{\rm B}d^5x{\sqrt{-g}}\Big ( {\cal R}+{1\over2}\big ({\nabla}\psi\big )^2\Big )\ +\ \int_{\rm BD} {\cal H}_3\wedge d\psi\ .\label{gauge-702}
\ee
An axionic coupling underlying a instanton has been argued to incorporate the quintessence effect \cite{priyabrat-EPJC}. We exploit the dynamical essence of an axionic scalar via its dual and construct Riemann type left (L) and right (R) duals in higher dimensional $d\ge5$ gravity theory. In particular, the Riemann type dual tensors are worked out in five dimensions using the the four-form ${\cal F}_{4}$ (as a generalized form of Levi-Civita tensor density) and they are:
\be
 \quad \qquad {\cal R}^{(L)}_{\mu\nu\lambda\rho}=\sqrt{{\pi\alpha'}\over2} {{\cal F}_{\mu\nu}}^{\alpha\beta}\ {\cal R}_{\alpha\beta\lambda\rho}\quad
{\rm and}\quad {\cal R}^{(R)}_{\mu\nu\lambda\rho}=\sqrt{{\pi\alpha'}\over2} 
{\cal R}_{\mu\nu\alpha\beta}\ {{\cal F}^{\alpha\beta}}_{\lambda\rho}\ .\label{gauge-704}
\ee
The generalized Riemann type L and R duals are new of its kind and may provide a clue to the quintessence axionic scalar dynamics. Interestingly, they reduce to the typical Riemann duals in $(3+1)$-dimensions as ${\cal F}_{\mu\nu\lambda\rho}\rightarrow {\cal E}_{\mu\nu\lambda\rho}$ there. 
Remarkably, the geometric duals reveal an intrinsic coupling of ${\cal F}_4$ with the Riemann type tensors in a NP-theory of emergent quantum gravity. The coupling ensures the presence of an axionic scalar in $d$$=$$5$ and hence signals the importance of a quintessence correction to the GTR. Quintessence is known to be a potential candidate to describe the dark energy in universe. Interestingly, the hidden quintessence (axionic scalar) dynamics to the GTR may be viewed as a NP-dynamical correction. 

\section{Emergent theories from superstrings}
\subsection{{\it Form} or {\it Fundamental}  theory}
\noindent
The emergent gravity in $(4+1)$-dimensions (\ref{gauge-70}) is governed by ${\cal H}_3$$=$$d^{\cal D}B_2^{(NS)}$ in $1.5$ order formulation with ${\cal D}_{\mu}$ as underlying covarinat derivative. The emergent metric is essentially governed by a (massive) NS field dynamics in first order within the purview of $d$$=$$10$ type IIA and IIB superstring theories. A massive NS field theory may be realized via Higgs Mechanism in an emergent gauge theory on a fat $3$-brane\cite{PTEP-NS}. 

\sp
\noindent
Furthermore $H_3$$=$$d^{\nabla}B_2^{(KR)}$ is sourced by a string charge and is best described on a $D_5$-brane due to a CFT. A $D_5$-brane is sourced by a 6-form in NS-NS and in RR sectors of type IIB superstring. It re-ensures a string/five-brane duality in $d$$=$$10$ superstring theory. Thus, the significance of a $3$-brane in an emergent gravity 
may demand a $d$$=$$12$ {\it fundamental} theory. Interestingly, F-theory has been viewed as a re-formulation of type IIB superstring theory with $D$-instanton as its fundamental unit \cite{vafa}. The $SL(2,Z)$ symmetry in type IIB supergravity has been exploited by the author to obtain a $d$$=$$12$ theory \cite{kar-NPB}. The field theoretic construction in $d$$=$$12$ leading to F-theory has widely been studied \cite{pope-CQG,ueno-JHEP}. 

\sp
\noindent
In fact, a dynamical ${\cal H}_3$ decouples to leave behind a torsion free geometry. Analysis leading to a stable gravitational $3$-brane as a fundamental unit in an emergent gravity provides a clue to the  {\it form} theory in $d$$=$$12$. 
In addition, an emergent gravity sourced by a KR field may formally be generalized to obtain a space filling emergent pair of $(8{\bar 8})$-brane underlying a $D_9$-brane in type IIB superstring. In principle, a $9$-brane would be governed by $F_{11}$$=$$dB_{10}$ and hence the $10$-form theory requires a minimum of $d$$=$$12$. 
The hint signifies the role of a constant $F_5$$=$$dB_4$ in an emergent gravity (\ref{gauge-70}). The constant turns out to be dynamical in six and higher dimensions. Importantly, a constant in $d$$=$$5$ describes a nontrivial topological coupling in addition to a total divergence $\left ({\cal F}_4\wedge {\cal F}_4\wedge {\cal F}_4\right )$ in $d$$=$$12$. We set $(2\pi\alpha'=1=\kappa')$ for all onward expressions and then the {\it fundamental} action is a priori given by
\be
S=\int d^{12}x \sqrt{-{\hat g}}\Big ( {\hat{\cal R}} -{1\over{48}}{\cal F}_4^2 - {1\over{240}} F_5^2\Big ) +\int B_4\wedge {\cal F}_4\wedge {\cal F}_4\ .\label{form-1}
\ee
where ${\cal F}_4$ governs a ${\cal H}_3$ potential. Interestingly an emergent gravitational $3$-brane is governed by the {\it form} theory in $d$$=$$12$. The Ricci type curvature scalar ${\hat{\cal R}}$ governs a NS field dynamics and is believed to describe an emergent metric (\ref{gauge-metric}) in {\it form} theory. Intuitively the (emergent) metric dynamics may primarily be viewed via the Riemann tensor and hence ${\hat{\cal R}}$ may be identified with a Ricci scalar. The 4-form signifies a NP-correction underlying a propagating ${\cal H}_3$. However, the $F_5$$=$$d^{\cal D}B_4$ may source a gravitational $3$-brane in {\it form} theory which may be re-expressed as:
\be
S=\int d^{12}x \sqrt{-{\hat g}}\Big ( {\hat{\cal K}}^{(S)} -{1\over{48}}{\cal F}_4^2 - {1\over{240}} F_5^2\Big ) + \int B_4\wedge {\cal F}_4\wedge {\cal F}_4\ .\label{form-2}
\ee
The gauge invarinat forms are given as
\begin{eqnarray}\label{forms-12d}
{\cal K}^{(s)}&=&-\frac{1}{4}\,{\cal H}_{3}^{2}\,\,\, , \quad {\rm where}\,\, \quad {\cal H}_{3}=d^{\cal D}{B}_{2}^{(NS)} \ .\nonumber
\end{eqnarray}
In addition
\begin{eqnarray} 
{\cal F}_{4}&=& d^{\cal D}{\cal H}_{3}\,\quad \qquad{\rm and}\qquad \quad {F}_{5}=d^{\cal D}{B}_{4}\ .
\end{eqnarray}
The first two terms in \eqref{form-2} possess their origin in an emergent NP-theory of gravity underlying a geometric torsion ${\cal H}_3$. The field 
strength $F_5$$=$$d^{\cal D}B_4$ possibly ensures an emergent gravitational $3$-brane. This is in agreement with the generalization of a fact that a point particle is sourced by an one-form gauge field, a string by two-form, a membrane by three-form, and a $3$-brane by $B_4$.

\subsection{Higher dimensional NP-theory}
\noindent
The total local degrees in {\it form} theory (\ref{form-2}) turns out to be $l$$=$$375$. It includes the NP-local degrees $l_{\rm NP}=120$ sourced by a dynamical geometric torsion and is given by ${\cal F}_4$, The remaining $255$ local degrees are sourced by the NS field with $l_{\rm NS}$$=$$45$ and a $4$-form field with $l_{\rm 4F}$$=$$210$. The value of $l_{\rm NP}$ decreases the dimension of space-time with $l_{\rm NP}$$=$$l_{\rm NS}$ in $d$$=$$7$ and the minimal NP degree $i.e.\ l_{\rm NP}$$=$$1$ is in $d$$=$$5$. 

\sp
\noindent
Generically, the NP degrees of freedom in $(d$$+$$1)$-dimensions is equal to the degrees of freedom of NP-theory underlying the NS-field and geometric torsion field in $d$-dimensions, $i.e.\ l_{\rm NP}^{d+1}$$=$$l_{\rm NS}^d$$+$$l_{\rm NP}^d$. Furthermore, $l_{\rm NS}^d$$=$$l_{\rm metric}^{d-1}$$+$$l_{\rm HS}$, where the HS (Higher-essence/dimensional Scalar) local $l_{\rm HS}$$=$$1$. The two level correspondences between a {\it form} theory in $(d+1)$-dimensions via a NP-theory in $d$-dimensions to a metric theory in $(d$$-$$1)$-dimensions is remarkable and is indeed a potential tool to describe an emergent NP theory on a gravitational pair of $(M{\bar M})$-brane\cite{Physica-NDS}. For instance, the $d$$=$$12$ (four) {\it form} theory can be mapped to $d$$=$$11$ non-perturbation ($M$) theory. One might lead to suppose that the NP-theory can further be viewed as a metric theory possibly corresponding to ten dimensional (space filling brane) superstring theories. Similarly a six dimensional form theory via a five dimensional (non-supersymmetric) NP-theory can be mapped to the GTR. Intuitively, it underlies an analogy between two sets: ($d$$=$$12$ {\it form} theory, $d$$=$$11$ M-theory and $d$$=$$10$ superstrings) and ($d$$=$$6$ form theory, $d$$=$$5$ NP-theory, the GTR).
The details of this analysis is beyond the scope of this paper.

\sp
\noindent
The {\it form} theory action on $S^1$ is worked out for the massless fields to describe a total $l$$=$$375$ local degrees on an emergent gravitational pair of $(M{\bar M})$-brane. Formally, the irreducible curvatures ${\hat{\cal K}}$, ${\cal F}_4$$=$$d^{\cal D}{\cal H}_3$ and $F_5$$=$$d^{\cal D}C_4$ (\ref{form-2}) are respectively reduced to a pair of curvatures $({\tilde{\cal K}}, {\cal F}_2^2)$, $({\cal F}_4, {\cal K})$ and $(F_4$$=$$d^{\cal D}{\cal B}_3, F_5)$ on $(M{\bar M})$-brane with ${\cal D}_{\mu}$ as the covariant derivative. The forms governed by these gauge invariant field strengths in terms of their respective underlying derivatives is given as
\begin{eqnarray}
\tilde{\cal K} &=&-\frac{1}{4}\,{\cal H}_{3}^{2}\quad ,  \quad {\rm where}\qquad {\cal H}_{3}=d^{\cal D}{B}_{2}^{(NS)}\nonumber \\
{\rm and}\qquad{\cal K}&=&-\frac{1}{4}\,{\cal B}_{3}^{2}\quad ,  \quad {\rm where}\qquad {\cal B}_{3}=d^{\cal D}{C}_{2}^{(NS)} \ .\nonumber
\end{eqnarray}
In addition
\begin{eqnarray} 
{F}_{2}&=& d^{\cal D}{A}_{1}\quad ,  \quad  {\cal F}_{4}=d^{\cal D}{\cal H}_{3}\quad , \quad {F}_{5}=d^{\cal D}{C}_{4}\quad \quad{\rm and}\quad{F}_{4}=d^{\cal D}{\cal B}_{3}\ .
\end{eqnarray}
The dimensionally reduced $d$$=$$11$ action a priori is given by
\bea
 &&S=\int_M d^{11}x {\sqrt{-{\tilde g}}}\ \Big ( {\tilde{\cal K}}\ -{1\over{48}}{\cal F}^a_4N_{ab}{\cal F}^b_4\Big )\ +\int_{\bar M} d^{11}x{\sqrt{-{\tilde g}}}\ \Big ({\cal K}\ -{1\over{4}} {\cal F}_2^2\ -{1\over{240}} F_5^2\ \Big )\nonumber\\
&&+\int_{(M{\bar M})}\Big [ {\cal H}_3\wedge \Big ({\cal F}_4\wedge {\cal F}_4+ {\cal F}_4\wedge F_4 + F_4\wedge F_4\Big )\ + F_2\wedge \Big ({\cal F}_4 +{\cal F}_4\Big )\wedge F_5\Big ]\label{form-3}
\eea
where $N_{ab}$ describes a $(2\times 2)$ diagonal matrix. In the action, the superscript $(S)$ (for symmetric) on the irreducible curvatures ${\tilde{\cal K}}$ and ${\hat{\cal K}}$ has been omitted and will be so in all the following equations.   
Generically, the second term on a $M$-brane ensures two $4$-forms: ${\cal F}_4$ and $F_4$. The matrix element $N_{11}$ ensures a non-canonical potential coupling only to the ${\cal F}_4$, which is sourced by the ${\cal F}_4$. Since ${\cal H}_3$ in $d$$=$$12$ does not couple to $F_5$, the element $N_{22}$ is a constant. The second term contains a NP-correction sourced by a propagating ${\cal H}_3$ in addition to $F_4$.

\sp
\noindent
Interestingly, the bulk {\it form} theory leads to a boundary description for an emergent $M$ brane on $S^{1}$. However, by invoking a generic bulk (NS-field)/boundary (metric tensor) correspondence\cite{Physica-NDS}  between the {\it form} theory and $M$-brane, the curvatures ${\hat{\cal K}}$, ${\cal F}_4$ and ${\cal F}_5$ in eq(\ref{form-2}) are respectively re-expressed as: $(R, \phi)$, $({\cal F}_4, {\cal K})$ and $(F_4, F_5)$ on an emergent pair of $(M{\bar M})$-brane, where $R$ and $\phi$ respectively denote the Ricci scalar and a (higher essence) scalar field. Needless to mention that $45$ local degrees of NS field underlying ${\hat{\cal K}}$ in $d$$=$$12$ are described in $d$$=$$11$ by $44$ local degrees of an emergent metric field and one local degree of a scalar field.  In fact, the $\phi$ field is an extra $12$-th transverse dimension inbetween a pair. 
Here also, the underlying derivative is ${\cal D}_{\mu}$ and the potentials governed by the gauge invariant forms are given respectively as
\begin{eqnarray}
{\cal F}_{4}&=&d^{\cal D}{\cal H}_{3}\quad  {\rm and} \quad {\cal K}=-\frac{1}{4}\,{\cal H}_{3}^{2}\quad , \quad {\rm where}\quad {\cal H}_{3}=d^{\cal D}{B}_{2}^{(NS)} \ ,
\end{eqnarray}
and
\begin{eqnarray}
{F}_{5}&=&d^{\cal D}{C}_{4}\qquad {\rm and} \qquad {F}_{4}=d^{\cal D}{H}_{3}\ .
\end{eqnarray}
The $d$$=$$11$ action (\ref{form-3}) reduces to yield: 
\bea
 \quad S&=&\int_M d^{11}x \ {\sqrt{-G}}\ \Big ( R_G\ -\ {1\over{48}}{\cal F}^a_4N_{ab}{\cal F}^b_4\Big )+\int \Big ({\cal H}_3\wedge{\cal F}_4\wedge {\cal F}_4\ +\ {\cal B}_3\wedge F_4\wedge F_4\ \Big )\nonumber\\
 &+&\int_{\bar M}\ d^{11}x{\sqrt{-G}}\ \Big ( {\cal K}\ - {1\over2}({\cal D}\phi)^2\ -{1\over{240}} F_5^2\Big )\ ,\label{form-4}
\eea
where $G_{\mu\nu}$, in the invariant-volume, denotes a dynamical (emergent) metric obtained under a bulk/boundary correspondence on a pair of $(M{\bar M})$-brane. The NP-action (\ref{form-4}) may a priori be identified with an eleven dimensional $M$-theory. However, a NP-correction is sourced only by a geometric torsion potential ${\cal H}_3$ in $1.5$ order formulation. Thus, the $M$-brane dynamics within a pair effectively takes account for the NP-dynamical correction and may formally be argued to govern the $M$-theory. Since, a gravitational pair is created across an event horizon of a background black hole in the NP-formulation \cite{abhishek-JHEP,abhishek-PRD}, the scalar field $\phi$ in $d$$=$$11$ is believed to play the role of a hidden or higher-essence in the disguise of an extra transverse dimension. 

\sp
\noindent
A varying thickness of the fat brane can be fixed by freezing the local degree of scalar field, $i.e.\ \phi\rightarrow \phi_0$. It disconnects an emergent $M$-brane in a pair from the ${\bar M}$-brane, where the radius $R$ of $S^{1}$ takes a fixed value $\phi_0$. The emergent curvatures and hence the causal effects on a gravitational $M$-brane universe are de-linked from that on the ${\bar M}$-brane as the $d$$=$$12$ coordinate system breaks down in the limit $\phi\rightarrow\phi_0$. Intuitively, the higher dimensional limit 
may imply that the polar angle $\theta\rightarrow 0$ on $S^{1}$. The apparent angular deficit angle $\theta$ under a Wick rotation may be viewed as light-like. Thus the disconnected emergent causal geometries are separated by a light-like cone. For an observer in a $M$-brane universe, a causal effect can be  re-interpreted as a spacelike event on ${\bar M}$-brane universe. The de-linked effective actions are given by 
\bea
\qquad S_M&=&\int d^{11}x{\sqrt{-G}}\ \Big ( R_G -{1\over{48}}{\cal F}^a_4N_{ab}{\cal F}^b_4\Big )+ \int \Big ({\cal H}_3\wedge{\cal F}_4\wedge {\cal F}_4+ {\cal B}_3\wedge F_4\wedge F_4\Big )\nonumber\\
{\rm and}\;\ S_{\bar M}&=&\int d^{11}{\bar x}\ {\sqrt{-{\bar G}}}\ \Big ( {\cal K}\ -{1\over{240}}F_5^2\Big )\ .\label{form-5}
\eea
In addition to the (bosonic) dynamics of $M$-theory (\ref{form-5}), a de-linked $M'$-theory in $d$$=$$11$ representing the ${\bar M}$-brane is remarkable. It is important to observe that the $M'$-theory is not a NP-theory as it does not describe a propagating geometric torison. 
Further investigation may reveal a plausible map if there is any between a $M'$-theory on $S^1$ and the $d$$=$$10$ type IIB (super)string theory.

\sp
\noindent
In a low energy limit, the NP-dynamical correction decouples and hence the limit is identified as a decoupling limit. The geometric torsion dynamics freezes in the limit ${\cal H}_3\rightarrow {\cal H}_3^0$ (a constant) and the $M$-brane action becomes
\be
S_M=\int d^{11}x{\sqrt{-G}}\ \Big (R_G -{1\over{48}}F_4^2\Big )+\int B_3\wedge F_4\wedge F_4\ .\label{form-6}
\ee
Interestingly, the decoupled emergent $M$-brane dynamics formally identifies with the bosonic sector of $d$$=$$11$ SUGRA \cite{cremmer-julia-scherk} in natural units. The presence of fermionic local degrees are not affected by the decoupling of a boson dynamics.  
Arguably, the absence of $\phi$-dynamics in a limit restores the supersymmetry and may lead to the SUGRA theory. The low energy limit of the $M$-brane is consistent with the $M$-theory in the same limit where $M$-theory is known to describe the $d$$=$$11$ SUGRA. This result suggests  that a geometric torsion is a plausible candidate to describe a complete NP-theory of quantum gravity and hence the torsion quanta may be interpreted as a ``graviton'' there. It may resolve some of the mystries related to the origin of dark energy in the  universe. A result that the GTR may be realized in a decoupling dynamics of a geometric torsion in superstring theory is new and is believed to enlighten the legacy of non-perturbation gravity to its depth.

\section{Concluding remarks}
\noindent
A non-perturbation theory of emergent gravity constructed by the author in collaborations \cite{abhishek-JHEP,abhishek-PRD,abhishek-NPB-P, sunita-NPB,priyabrat-EPJC} had been revisited to show the significance of a four-form ${\cal F}_4$ correction to the torsion free geometries in Einstein gravity. Interestingly, the quantum correction had been argued to be sourced by a instanton, underlying an axionic scalar, in a string-brane setup,  which in turn may be coined as  a ``fat'' brane.  Unlike a non-gravitational or flat $D$-brane, a fat-brane absorbs the  local degree of freedom of an axionic scalar and takes account for the space-time curvature by coupling to the closed string modes. The Riemann duals were shown to absorb a dynamical axion in a semi-classical description and is argued to describe the quintessential cosmology. Importantly, a higher dimensional generalization was obtained in eqs.(\ref{form-1})-(\ref{form-2}) and they hint for a {\it fundamental} or {\it form} theory in $d$$=$$12$. Further analysis has revealed an emergent gravitational pair of $(M{\bar M})$-brane from the {\it form} theory. In a semi-classical limit, the $M$-brane dynamics has formally been identified with the $d$$=$$11$ SUGRA. It is believed to provide a wide scope to enhance the horizon underlying certain aspects of dark energy and may unfold a clue to model the dark gravity.

%%%%%%%%%%%%%%%%%%%%%%%%%%%%%%%%%%%%%%%%%%%%%
%\section*{Acknowledgements}
%Author gratefully thanks Josheph Polchinski, Fernando Quevedo and John H. Schwarz for insightful discussions at an early stage of the research progress. 

%**********************************************************%
%\section*{Refrences}
\def\anp{Ann. Phys.}
\def\cmp{Comm. Math. Phys.\ {}} {}
\def\springer{Springer. Proc. Phys.}
\def\prl{Phys. Rev. Lett.}
\def\prd#1{{Phys. Rev.} {\bf D#1}}
\def\jhep{JHEP\ {}}{}
\def\cqg{Class.\& Quant.Grav.}
\def\plb#1{{Phys. Lett.} {\bf B#1}}
\def\npb#1{{Nucl. Phys.} {\bf B#1}}
\def\mpl#1{{Mod. Phys. Lett} {\bf A#1}}
\def\ijmpa#1{{Int. J. Mod. Phys.} {\bf A#1}}
\def\ijmpd#1{{Int. J. Mod. Phys.} {\bf D#1}}
\def\mpla#1{{Mod. Phys. Lett.} {\bf A#1}}
\def\rmp#1{{Rev. Mod. Phys.} {\bf 68#1}}
\def\jaat{J.Astrophys.Aerosp.Technol.\ {}} {}
\def \epj#1{{Eur.Phys.J.} {\bf C#1}} 
\def \jcap{JCAP\ {}}{}
\def\physica{Physica Scripta }
\def\ptp{ Prog.Theo.Exp.Phys.}
\def\ijtp{Int. J. Theo. Phys.}
%**********************************************************%

\end{document}